\documentstyle[11pt,newpasp,twoside,epsf]{article}
\def\edcomment#1{\iffalse\marginpar{\raggedright\sl#1\/}\else\relax\fi}
\reversemarginpar

\begin{document}
\title{Indirect searches for SUSY Dark Matter with the MAGIC Cherenkov Telescope}
\author{J. Flix(1), M. Martinez(1) and F. Prada(2)(3) for the MAGIC Collaboration (*)}
\affil{(1) Institut de F\'{\i}sica d'Altes Energies, Edifici Cn, Universitat Autonoma de
Barcelona, Cerdanyola del Vall\'{e}s, 08193, Spain \\
(2) Instituto de Astrofisica de Canarias, Canary Islands, Spain. \\
(3) Instituto de Astrofisica de Andalucia-CSIC, Granada, 18008, Spain \\
(*) Updated collaborator list at: http://www.magic.iac.es
}

\begin{abstract}
Neutralinos are the natural well-motivated candidates to provide
the non-baryonic dark matter of the universe which may produce
detectable signals through their annihilation into neutrinos,
photons or positrons. Due to its high flux sensitivity and low
energy threshold, the MAGIC Cherenkov telescope could potentially
detect the neutralino annihilation high energy photon products. In 
the framework of minimal supergravity, the neutralino
SUSY parameter space can be scanned in different benchmark
scenarios defined after accelerator and cosmology constraints.
Moreover, the neutralino density profiles in galaxy halos and
sub-halos have to be understood to infer which is the optimal
observation region to be explored by the MAGIC telescope for the
detection of neutralino photon signatures within our Galaxy and
our Local Group.
\end{abstract}

\section{Introduction}

Observational cosmology is showing us a universe which is almost
dominated by dark matter. Revealing its nature constitutes one of
the most challenging questions of modern cosmology and particle
physics, both from the theoretical and the experimental sides. A
Cold Dark Matter scenario (CDM) in which the non/baryonic matter
contribute to approximately 27$\%$ of the universe energy density
content is presently widely accepted. Weakly Interacting Massive
Particles (WIMPs), non-relativistic relics from the early
universe, are good candidates to account for dark matter. Any of
the known particles have to satisfy the requirements of being
weakly interacting, massive and stable enough to be a relic
particle. Supersymmetric (SUSY) extensions of the Standard Model
of particle physics provide a particle that can accommodate all
the requirements: the lightest SUSY particle (LSP), which turns
out to be the neutralino in most of the SUSY-breaking scenarios.

Besides instruments based on the detection of the energy deposited
by elastic nucleus-WIMP scattering in underground massive
bolometers, others take profit of the fact that the neutralino is
a Majorana particle and, therefore it can pair-annihilate
producing high-energy neutrinos, positrons, antiprotons and
gamma-rays (Bergstrom et al. 2000). Unfortunately, so far there is
no firm evidence by any of these techniques of a dark matter
particle detection. Nevertheless, currently a new door is being
open with new detectors which might be able to find the WIMPs. We
will concentrate on the discussion of the MAGIC telescope
potential: an instrument of a new generation of ground-based
Imaging Air Cherenkov Telescopes (IACT), which might be able to
detect gamma-rays from neutralino annihilations.

\section{Spectral signatures of $\chi$ annihilations in Dark Matter halos}

Neutralinos pair-annihilate through a variety of channels but for
the present discussion only those that generate gamma-radiation in
some intermediate or final state are of our interest. Annihilation
processes like $\chi\chi\rightarrow\gamma\gamma$ and
$\chi\chi\rightarrow\gamma Z$  generate spectral gamma lines. A
distinct process is $\chi\chi\rightarrow q \bar{q}$, which
generates a continuum of gamma-rays mainly by the decay of
$\pi^{0}$-mesons produced within these jets.

Given a neutralino density profile  $\rho_{CDM}(r)$ (i.e. the dark matter density profile
within a galaxy), the expected differential gamma-ray flux from its
annihilation along the observation line of sight is given by:

\begin{eqnarray}
\frac{\Phi (E_{\gamma})}{dE_{\gamma}} = & [ N_{\gamma\gamma}b_{\gamma\gamma}\delta(E_{\gamma} -
                     M_{\chi}) +  N_{\gamma Z}b_{\gamma Z}\delta(E_{\gamma} -
                     M_{\chi}(1-\frac{M_{Z}^{2}}{4M_{\chi}^{2}})) \nonumber \\
                   & + \displaystyle{ \sum_{F}} \frac{dN_{\gamma}^{F}}{dE_{\gamma}}b_{F}] \frac{<\sigma v>}{2 M_{\chi}^{2}}
                     \frac{1}{4\pi} \int_{l.o.s} \rho_{CDM}(r) dr
\end{eqnarray}

where $N_{i}$ is the number of photons produced per annihilation
channel (2 for $\gamma\gamma$, 1 for $\gamma Z$ and
$n(E_{\gamma})$ for the case of the continuum running over all the
F final states), $b_{i}$ are the process branching ratios,
$<\sigma v>$ is the averaged product of total annihilation cross-section
and relative velocities and $M_{\chi}$ is the neutralino mass. One
should average all the flux contributions over the detector solid
angle.

\subsection{Modeling neutralino properties}

In order to model the neutralino properties, a few SUSY scenarios
can be considered. We restrict our attention to a constrained
version of the Minimal Supersymmetric Standard Model (CMSSM),
which incorporates a minimal supergravity scenario (mSUGRA) of
soft supersymmetry breaking. In this context, given a set of input
SUSY parameters at the Grand Unification scale, the whole spectrum
of SUSY particles can be derived to low energies from the
renormalization group equations of these parameters. This includes
the neutralino properties as mass, annihilation cross-sections,
branching ratios and particle spectrum from annihilations.

The precise electroweak data from LEP suggest that the Higgs boson should be
very light, in good agreement with the prediction of MSSM models. Very
accurate measurements of the gauge couplings are consistent with a
supersymmetric Grand Unified Theory, if supersymmetric partners of the
standard model particles weight less than about 1 TeV. Neutralino relic
abundance falls naturally within the WMAP favored range, below this energy.

Several benchmark models to set the CMSSM input parameters were initially
proposed to provide a common way to compare the SUSY discovery potential of
future accelerators as LHC. The benchmark points have been recently re-defined
after the very precise WMAP results. They have been chosen to fulfill conditions
imposed by LEP measurements, $b\rightarrow s\gamma$ and the current relic
density range $\Omega_{CDM}$. With
these constraints, the effective dimensionality of the CMSSM parameter space
has been reduced (Battaglia et al. 2003). Indeed it is quite more reduced if
the constraint from the anomaly on $g_{\mu}-2$ measurement is also included in the analysis.

The most important direct experimental constraints on the MSSM parameter space
are provided by LEP searches of SUSY particles and Higgs Bosons. Based on
data-taking up to center-of-mass energy of 208 GeV, a lower limit can be set
to the neutralino mass: of about 108 GeV.  As this lower limit depends on
calculation of the Higgs mass lower limit and the top mass measurement, a
conservative lower bound would become to about 85 GeV, if one takes
theoretical and measurement uncertainties into account (Battaglia et al. 2001). An upper limit on the
neutralino mass can be set from the upper limit on the
high accurate current $\Omega_{CDM}$ measurement, which imposes that $M_{\chi}$ is of about
400 to 500 GeV under mSUGRA (Ellis et al. 2003).

\subsection{Modelling the Dark matter density profiles}

Due to the small thermal velocity of CDM particles, their density
fluctuations survive from the early universe on all scales. In
this framework, N-boby simulations have shown that structures
develop as small clumps collapse, undergoing series of merging
that results in hierarchical formation of massive galactic dark
matter halos, as the universe expands. The most massive halos are
the hosts for baryonic systems such as Galaxies. According to this
hierarchical clustering scenario, numerous dark matter satellites
(sub-halos) of different sizes and masses are revealed to orbit
the virialized massive dark matter halos (Klypin et al. 1999).

The CDM scenario predicts a larger number of sub-halos in
galactic halos, more than known dwarf satellites of the
Milky Way. This could be possible if there would be a large number
of sub-halos being the known compact high-velocity
clouds (CHVCs), which may possibly trace them. If neutralino is the dark matter
candidate, it is interesting to correlate the locations of
high-velocity clouds with the unidentified EGRET gamma-ray
sources, and estimate how much Milky Way sub-halos potentially
would emit gamma-rays by neutralino annihilation, which would have
been detected by the satellite (Flix et al. 2003).

It is clear from (1) that the gamma-ray fluxes should be maximal
for the closest densest dark matter regions of the universe, i.e,
one would expect higher fluxes for cores of galaxies and
sub-halos. How the density profile behaves at the very center is
widely under discussion as more high-resolution numerical
simulations are available. This is really fundamental because it
affects directly the gamma-ray flux predictions.

Including the adiabatic compression of the dark matter due to baryonic infall
during galaxy formation results in an increase of the central density of dark
matter. This effect should be taken into account as an enhancement of the
annihilation fluxes is expected (Prada et al. 2003).

\section{Detection of dark matter with MAGIC}

The 17 m diameter f/D=1 MAGIC telescope (Barrio et al. 1998) is
the largest of the new generation of Imaging Atmospheric Cherenkov
Telescopes. MAGIC is located at the Canarian island of La Palma
(28.8 N, 17.9 W) at the Roque de los Muchachos observatory (ORM),
2200 m above sea level. It detects gamma-rays in the high energy
regime with an energy threshold of about 30 GeV, providing much
larger effective areas (and much superior sensitivity, of about
10$^{-10}$ to 10$^{-11}$ cm$^{-2}$s$^{-1}$ for E = 30 GeV) than
satellite detectors, good angular resolution, acceptable energy
resolution and a well tested capability to separate gammas from
backgrounds. The MAGIC telescope is in its final commissioning
phase and it is expected to start regular observations by the end
of this year.

Apart from a wide variety of astrophysical sources of interest,
like AGNs, pulsars, GRBs or SNRs, MAGIC is a suitable instrument
for dark matter searches for neutralino annihilation signals from
the Galactic Center and Local Group galaxies.

The Galactic Center is a crowded region containing sources
that can emit also in gamma-rays (as a SNR, the most brilliant
unidentified EGRET sources, ...). If diffuse gamma fluxes from
annihilations are strong enough, as suggested by adiabatic
compression models, observations displaced from the Galactic
Center would be preferred. Also, we notice that the energy
threshold of a Cherenkov telescope increases with the zenith angle
$\Theta$ of the observation (this results in an energy threshold of 200
GeV for MAGIC in observing the Galactic Center). Because the very
sharp energy spectral features, low zenith angle observations for
Dark Matter searches are preferred. It is very attractive indeed
that the effective areas are bigger, so that an improvement of the
sensitivity is achieved.

\section{Summary}

Extensive montecarlo understanding of he detector response at
different zenith angles is being done as well as taking into
account the most important background sources as the cosmic
proton, electron and Helium fluxes, the galactic diffuse gamma
radiation as well as the expected annihilation signal one may
expect in different scenarios. This will allow us to select which
region is preferable for observations related to Dark Matter
searches within the Milky Way halo.

\end{document}